\documentclass[12pt,preprint]{aastex}

\usepackage{CJK}
\usepackage{natbib}
\usepackage{graphicx}

\begin{document}
\begin{CJK}{UTF8}{gbsn}

\title { EXTRASOLAR REFRACTORY-DOMINATED  PLANETESIMALS: AN ASSESSMENT}

\author{M. Jura\altaffilmark{a} \& S. Xu(许\CJKfamily{bsmi}偲\CJKfamily{gbsn}艺)\altaffilmark{a}}

\altaffiltext{a}{Department of Physics and Astronomy, University of California, Los Angeles CA 90095-1562; jura, sxu@astro.ucla.edu}

\begin{abstract}

Previously published observations of 60 externally-polluted white dwarfs show that  none of the stars have accreted from intact refractory-dominated parent bodies  composed mainly of Al, Ca and O, although planetesimals with such a distinctive composition have been predicted to form.   We propose that such remarkable objects are not detected, by themselves, because, unless they are scattered outward from their initial orbit, they are engulfed and destroyed during the star's  Asymptotic Giant Branch evolution. As-yet, there is at most only  weak evidence supporting a scenario where the composition of any extrasolar minor planet can be explained
by blending of an outwardly scattered refractory-dominated planetesimal with an ambient asteroid.  
  \end{abstract}
\section{INTRODUCTION}

How do rocky planets form?  Is Earth normal? Are there  extrasolar planetesimals with exotic compositions?  Although such questions have received an enormous amount of theoretical attention  \citep{Armitage2010}, there are relatively few observational constraints.  Studies of externally-polluted white dwarfs provide an opportunity to
address these topics.

Amassed evidence  \citep{Farihi2009, Farihi2010,Gaensicke2006, Gaensicke2007, Gaensicke2008, Jura2008, Kilic2006, Reach2005, Zuckerman2010} has provided compelling support for the scenario that the heavy elements in white dwarfs with $T$ $<$ 20,000 K usually derive from rocky planetesimals.  The widely accepted  model is that an asteroid's orbit is perturbed so that it  passes within the white dwarf's tidal radius where it is destroyed, a circumstellar disk is formed, and
eventually this material is accreted onto the host star \citep{Bonsor2011, Debes2002, Jura2003, Rafikov2011}.   Consequently, spectroscopic determination of the atmospheric abundances in externally-polluted white dwarfs is a  powerful
tool to measure the elemental compositions of extrasolar minor planets \citep{Zuckerman2007}.  This topic was comprehensively discussed in \citet{Jura2008} and more recently reviewed
in \citet{Jura2013}.

Recent observational progress in the study of externally-polluted white dwarfs has been substantial.  Largely by using the {\it Spitzer Space Telescope}, 30 white dwarfs with dust disks that lie within  the star's tidal radius have been identified \citep{Farihi2009, Farihi2012a, Brinkworth2012, Xu2012}.  The first measurement of all major elements in a polluted
white dwarf was performed by \citet{Klein2010}, and, as reported in more detail below, results for more stars have been announced.  Characteristic minimum
masses of the accreted parent bodies of the best studied systems require asteroids with diameters greater than 100 km, likely remnant planetesimals from
the era of planet formation.  These  data for externally-polluted white dwarfs therefore uniquely enable detailed assessment of models for the formation and evolution
of rocky planets.

 In conventional  planet formation models, the condensation of gas into solids is controlled by the gas temperature, pressure and composition in 
 the protoplanetary disk midplane.  For example, Earth is largely composed of material that  condenses at ${\sim}$1100 K \citep{Allegre2001}; it is presumed that this result reflects the local disk physical conditions at ${\sim}$1 AU from the Sun at the time when the first planetesimals that grew into Earth were formed \citep{Cassen1994,Hersant2001}.  Similarly,
 the existence of icy bodies in the outer solar system is explained by their formation beyond  a snow line, the region where  H$_{2}$O condenses  into ice and contributes as much as 50\% to the mass of a newly-formed
 planetesimal \citep{Lecar2006,Garaud2007,Oka2011}.

  There is now evidence from externally-polluted white dwarfs that these familiar models for thermal segregation in the solar system's protoplanetary disk  can be extrapolated to extrasolar planetary systems to explain  the measured volatile-element compositions of extrasolar planetesimals.  That is, formation interior to a snow line can explain the results of
 \citet{Dufour2012}  that ice comprised less than 1\% of the mass of the parent body accreted
 onto J073842.56+183509.06 and \citet{Jura2012b}  that H$_{2}$O was less than 1\% of the mass accreted
 onto an ensemble of nearby externally-polluted white dwarfs.  Additionally, carbon within extrasolar planetesimals is often deficient by at least a factor of ten  compared to its ``cosmic" abundance \citep{Jura2006,Jura2012a,Gaensicke2012,Koester2012}.  Therefore interstellar grains containing carbon that entered  the inner protoplanetary disks  must
 have  been vaporized where  gas temperatures   likely  exceeded 500 K \citep{Lee2010}.
 
 \citet{Bond2010a} have considered standard models for protoplanetary disks and proposed that  there is a zone with a temperature near 1400 K where only the most refractory elements condense to form long-lived planetesimals.  Local remnants of such a  process might be the calcium aluminum inclusions (CAIs) in meteorites that may have been produced in the inner solar system
  \citep{Ciesla2010} when the disk was
 no older than 3 ${\times}$ 10$^{5}$ yr \citep{Young2005} and perhaps as young as 2 ${\times}$ 10$^{4}$ yr \citep{Thrane2006}.    \citet{Sunshine2008a,Sunshine2008b} have proposed that some  main belt asteroids with high fractions of refractory elements  may be composed of  blends of exceptionally refractory planetesimals that were formed in the inner solar system and scattered
 outwards to collide and partially merge with an ambient object.  Although the magnitude of such refractory-rich contamination is uncertain \citep{Hezel2008}, perhaps similar radial mixing measurably occurred in extrasolar planetary systems.

 There are at least two ways by which we might infer the long-lived existence of  refractory-dominated planetesimals.  First, most obviously, such an object might be accreted intact by a white dwarf and the heavy elements in the atmosphere
 of the star would be largely Al, Ca and O.  Second, more subtly,  accreted  material might be a blend of  a more ``normal" Earth-like composition with  a  refractory-dominated  planetesimal  that was scattered outwards from its original orbit \citep{Carter-Bond2012}.   This blending may occur either by the collision of two asteroids and a subsequent partial or complete merging, or simply by two different
 asteroids being independently tidally-disrupted, but from our perspective, being simultaneously accreted by an externally-polluted white dwarf.
  
 In Section 2, we summarize relevant  observations of externally-polluted white dwarfs. In Section 3, we describe a toy model to explain why intact refractory-dominated planetesimals have
 not yet been found.  In Section 4, we consider the possibility that refractory-rich material in some observed polluted white dwarfs
 results from blending of refractory-dominated planetesimals into the observed contaminating material.  In Section 5 we
 put our results in perspective and in Section 6 we present our conclusions.  
 
 \section{CURRENT OBSERVATIONS}

 Ca II 3933 {\AA} is  the strongest heavy-element line in the optical spectrum of polluted white dwarfs cooler
 than 15,000 K \citep{Sion1990} while  Mg II 4481 {\AA} and/or Mg I between 3832 {\AA} and 3838 {\AA} often  also are detected.  Because refractory-dominated  planetesimals are predicted to be largely composed of O, Ca and Al, we use  
 observed value of $n$(Mg)/$n$(Ca) to determine if the white dwarf has accreted such a parent body.
Previously published results for 60 polluted white dwarfs\footnote{ Included are  26 stars from \citet{Koester2011},   8 stars  from \citet{Kawka2011} and 5 stars from \citet{Zuckerman2003}.   
Also included are  GD 40 and G241-6 \citep{Jura2012a}, GD 362 \citep{Zuckerman2007},  G149-28 and NLTT 43806 \citep{Zuckerman2011}, PG 1225-079 and HS 2253+8023 \citep{Klein2011}, NLTT 1675 and NLTT 6390 \citep{Kawka2012},  WD 0738-172 \citep{Koester2000},  van Manaan 2 \citep{Wolff2002},  PG 1015+161 and WD 1226+110 \citep{Gaensicke2012}, GALEX 193156.8+011745 \citep{Melis2011,Vennes2011}, J073842.56+183509.06 \citep{Dufour2012}, HE 1349-2305 \citep{Melis2012},  Ton 345 \citep{Gaensicke2008}, SDSS J095904.69-020047.6 \citep{Farihi2012a}, GD 61 \citep{Farihi2011a}, G77-50 \citep{Farihi2011b},  and GD 16 \citep{Koester2005}.}  
with expected relative errors less than 0.5 dex are displayed in Figure 1 where $n$(Mg)/$n$(Ca)  is plotted compared to the white dwarf's effective temperature, a measure of the
star's cooling age.  We also display  $n$(Mg)/$n$(Ca) for a representative type A CAI, a value that  is midway between the higher ratio measured for type B CAIs \citep{Grossman1980}
and the  lower ratio predicted for refractory-dominated planetesimals \citep{Bond2010a}.  We see that  
Mg is always more abundant than Ca, as  in bulk Earth but not in CAIs.  There is no evidence
for intact extrasolar  refractory-dominated planetesimals.

The average values of $\log$ $n$(Mg)/$n$(Ca) and $\log$ $n$(Fe)/$n$(Ca) for the ensembles of stars plotted in Figures 1 and 2 are 1.26 and 1.00, respectively.  The corresponding solar system values of these two ratios are 1.21 and 1.13, respectively \citep{Lodders2003}.  At least on average, the relative
abundances of Fe, Mg and Ca among extrasolar planetesimals are approximately solar system-like.  

\begin{figure}
 \plotone{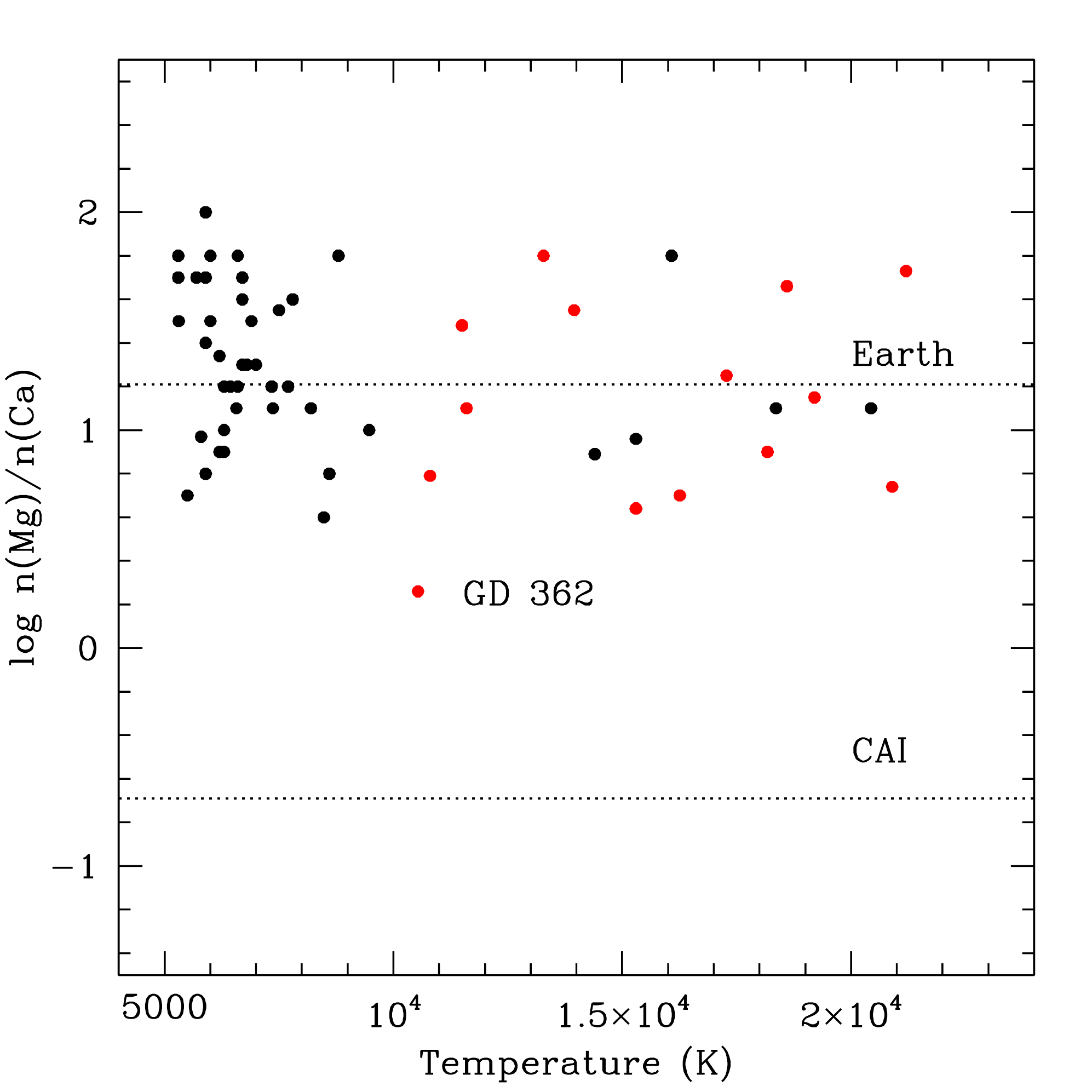}
\caption{Plot of published values of $\log$ $n$(Mg)/$n$(Ca) for 60 externally-polluted white dwarfs vs.
 stellar effective temperature.  Although each study is different, the typical errors bars, when reported, are 0.2-0.3 dex, and for clarity, they are suppressed here.
   The horizontal dashed lines are for bulk Earth \citep{Allegre2001} and a representative type  A CAI \citep{Grossman1980}. There is no system where $n$(Mg)/$n$(Ca) is 
close to the ratio found in CAIs.  Red and black points denotes white dwarfs with and without dust disks, respectively \citep{Farihi2009,Xu2012}.  There is no evidence that the accretion from stars with dust is compositionally different from stars without dust. }
\end{figure}

Because different elements gravitationally settle with different rates, the ratio of Mg to Ca in the star's photosphere need not measure the element ratio in
the accreted parent body \citep{Dupuis1993,Koester2009}.  The display in Figure 1 is for the ``instantaneous" approximation where the observed abundance ratio
directly measures the abundance ratio in the parent body.  If, instead, the system is in a steady state where the rate of accretion onto the top of the photosphere is balanced by
the loss rate at the bottom of the mixing zone, then the observed value of $n$(Mg)/$n$(Ca) must be altered by the ratio of the gravitational settling times to infer
the true ratio in the accreted asteroid.  Typically, this correction is between 0.1 dex and 0.2 dex \citep {Koester2009,Koester2011} and therefore does not affect
our qualitative conclusion.  If the system is in a decaying phase where there accretion is negligible and there is only gravitational settling, then the observed
value of $n$(Mg)/$n$(Ca) can become much larger than the value in the parent body because Mg typically settles more slowly than Ca \citep{Koester2009}.  However,
in this decaying phase we would also expect $n$(Fe)/$n$(Ca) to become quite small because Fe settles more rapidly than Ca.
Of the 60 stars for which data are shown in Figure 1, 50 also have measured values of the Fe abundance.   In Figure 2,  we display $n$(Fe)/$n$(Ca).  Again, there is no
evidence for refractory-dominated planetesimals.  

\begin{figure}
 \plotone{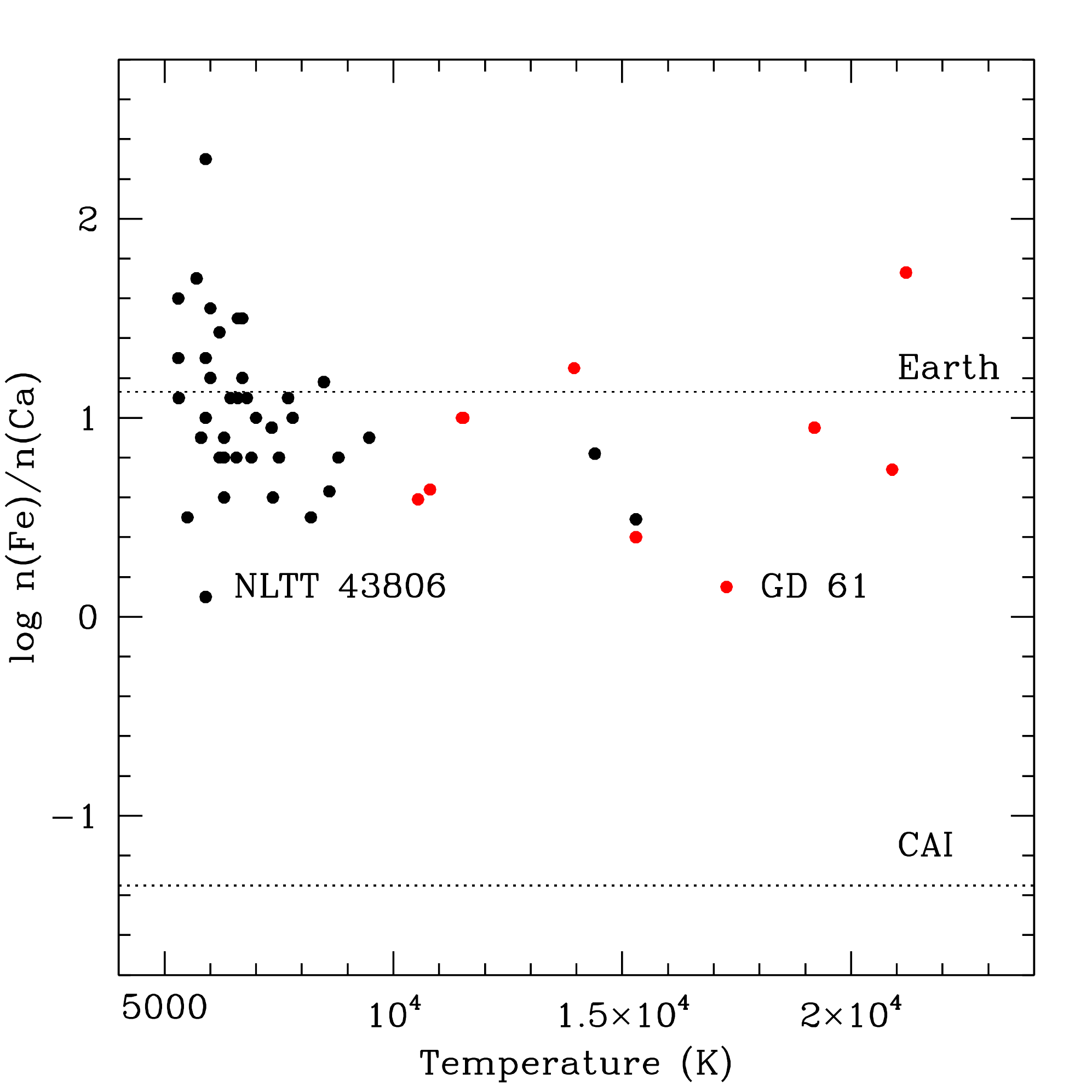}
 \caption{Same as Figure 1 except for $\log$ $n$(Fe)/$n$(Ca) for the 50 stars where Fe is reported.    There is no system where $n$(Fe)/$n$(Ca) is
close to the ratio found in CAIs.}
 \end{figure}

\section{DETECTABILITY OF  NON-SCATTERED REFRACTORY-DOMINATED PLANETESIMALS}

We now consider a very simplified model to assess the circumstances where we might be able to detect  intact refractory-dominated
planetesimals accreting onto white dwarfs.   We first estimate the orbital radius where such planetesimals may form in a protoplanetary disk.   We then
determine if the surviving  planetesimals are engulfed during the evolutionary phase when the star is on the Asymptotic Giant
Branch (AGB).  If not subsumed, we assume planetesimal survival into the phase when  the star is a white dwarf.   
In this section, we consider the case where the asteroid remains in the orbital location where it formed; in Section 4, we consider the possibility that some minor planets may be scattered outwards. 

\subsection{Formation Location}

To estimate the orbital location for the formation of refractory-domianted planetesimals, we extend the models of \citet{Bond2010a,Bond2010b} to higher stellar masses; they mainly considered planet formation around stars of about 1 M$_{\odot}$.  Therefore, we assume that 
refractory-dominated planetesimals  form where the midplane disk temperature, $T_{mid}$, is low enough that calcium and aluminum both form solid oxides,
yet high enough that silicon, magnesium and iron remain in the gas phase.  For the gas pressures of interest and for a solar system-like composition, this
requires that 1340 K ${\leq}$ $T_{mid}$ ${\leq}$ 1520 K for 50\% condensation \citep{Lodders2003}.  For simplicity, we focus on
a single condensation temperature  of 1400 K.

For a pre-main-sequence star of mass, $M_{*}$, which possesses a protoplanetary disk where material is accreting onto the host star with rate, ${\dot M_{*}}$, the effective
temperature at the surface of an opaque disk, $T_{e}$, at orbital distance, $R$, is controlled by  dissipation of
accretion energy.  For $R$ $>>$ $R_{*}$, where $R_{*}$ is the stellar radius, then \citep{Hartmann2009}:
\begin{equation}
T_{e}\;{\approx}\;\left(\frac{3\,G\,M_{*}\,{\dot M_{*}}}{8\,{\pi}\,R^{3}\,{\sigma}_{SB}}\right)^{1/4}
\end{equation}
Here, $G$ and ${\sigma}_{SB}$ denote the gravitational constant and Stefan-Boltzmann constant, respectively.
The midplane temperature is controlled by the vertical diffusion of radiation and is generally higher than the effective temperature at the disk surface.   If the disk has a total gas surface density of  ${\Sigma}_{H}$,
and if ${\chi}$ denotes the Rosseland mean opacity, then we define the vertical optical depth from the midplane through the dust layer as:
\begin{equation}
{\tau}\;=\;\frac{{\chi}\,{\Sigma}_{H}}{2}
\end{equation}
Because energy flows in both directions and there is no net flux at the midplane, the computed temperature as a function of optical depth is somewhat different from the thermal profile in a stellar atmosphere \citep{Hubeny1990, Oka2011}, and:
\begin{equation}
T_{mid}\;{\approx}\;\left(\frac{3\,{\tau}}{8}\right)^{1/4}\,T_{e}
\end{equation}

We now  estimate the disk surface density.  \citet{Bond2010b} presume a minimum mass solar nebula
where ${\Sigma_{H}}$ varies as $R^{-1.5}$.  However, because we wish to extrapolate from the solar system to other environments,
we adopt a simplified conventional steady state ${\alpha}$-disk configuration where viscosity at the midplane controls
the energy dissipation and the inward transport of mass\footnote{Much more sophisticated models have been presented \citep{Zhu2010}; including, for example,  the possibility of
a ``dead zone".}.  In this  case, then \citep{Oka2011}:
\begin{equation}
{\Sigma}_{H}\;=\;\frac{{\dot M_{*}}\,{\mu}}{3\,{\pi}\,{\alpha}\,k_{B}\,T_{mid}}\,\left(\frac{G\,M_{*}}{R^{3}}\right)^{1/2}
\end{equation}
where ${\alpha}$ is the usual dimensionless viscosity scaling factor\footnote{${\alpha}$ can be defined either in terms of the isentropic sound velocity  or the isothermal sound velocity, a difference of a factor of ${\gamma}^{1/2}$  where ${\gamma}$ is the ratio of specific heats.  Following \citet{Armitage2010} in recognizing that uncertainties in the viscosity dwarf this issue, we simply define ${\alpha}$ in terms of the isothermal sound velocity.} and ${\mu}$ is the mean molecular weight of the gas.  For our fiducial model, we adopt ${\alpha}$ = 0.01, but the uncertainty is substantial \citep{King2007}.

If the material condenses at orbital radius, $R_{cond}$, where the midplane temperature is $T_{cond}$, then from Equations (1) - (4), we find:
\begin{equation}
R_{cond}\;=\;\left(\frac{3\,{\chi}\,{\mu}}{128\,{\pi}^{2}\,{\alpha}\,k_{B}\,{\sigma}_{SB}\,T_{cond}^{5}}\right)^{2/9}\,(G\,M_{*})^{1/3}\,{\dot M_{*}}^{4/9}
\end{equation}

To determine $R_{cond}$, we  now
estimate  the   opacity  which depends upon the grain size distribution,  maximum particle size and composition.  Because we consider the regime where  refractory-dominated planetesimals form, we assume that grains are only composed of  minerals that are combinations of CaO and Al$_{2}$O$_{3}$.  If we assume that all of the available calcium and aluminum condense, then 
with abundances from \citet{Lodders2003} which are representative of G-type stars in the solar neighborhood \citep{Reddy2003}, the mass fraction compared to hydrogen that can be condensed into solids, $f_{cond}$, is 2.5 ${\times}$ 10$^{-4}$ \citep{Ciesla2010}.    
Although the grain size distribution is unknown, we are considering an environment where particles are growing into
planetesimals.  Therefore, we expect particles to be larger than interstellar grains.  It is possible that the grain size distribution follows a power law of the form \citep{D'Alessio2001}:
\begin{equation}
n(a)\,da\;{\propto}\;a^{-2.5}\,da
\end{equation}
If so, then most of the particles are large enough that the  Rosseland mean opacity is determined by  the particles' geometric cross sections.  If the maximum particle size is $a_{max}$, then:
\begin{equation}
{\chi}\;=\;f_{cond}\,\left({\int}_{0}^{a_{max}}\,{\pi}\,a^{2}\,n(a)\,da\right)\,\left({\int}_{0}^{a_{max}}\frac{4{\pi}a^{3}\,{\rho}_{s}}{3}\,n(a)\,da\right)^{-1}\;=\;\frac{9\,f_{cond}}{4\,{\rho}_{s}\,a_{max}}
\end{equation}
Although different minerals have different densities; for our fiducial model, we adopt 
${\rho}_{s}$ ${\approx}$ 3 g cm$^{-3}$.  If $a_{\max}$ = 1 mm \citep{D'Alessio2001}, then ${\chi}$ = 2.0 ${\times}$ 10$^{-3}$ cm$^{2}$ g$^{-1}$.  However, this opacity is uncertain by at least a factor of 10.

We now evaluate Equation (5) for our fiducial model describing the formation of refractory-dominated planetesimals.  If $M_{*}$ = 1.0 M$_{\odot}$, ${\dot M_{*}}$ = 3 ${\times}$ 10$^{-6}$ M$_{\odot}$ yr$^{-1}$, $T_{cond}$ = 1400 K and the gas is composed of molecular hydrogen and helium with a solar fractional abundance so that ${\mu}$ = 2.4 $m_{H}$ where $m_{H}$ is the mass of a hydrogen atom, we expect $R_{cond}$ = 0.38 AU.   Therefore, for this stellar mass and accretion rate, we reproduce the results 
computed in much more detailed models \citep{Bond2010a}.  

The temperature in the disk midplane being as high as ${\sim}$1400 K is  mainly a consequence of
the accretion rate being greater than 10$^{-6}$ M$_{\odot}$ yr$^{-1}$ \citep{Hersant2001,D'Alessio2005}.
This presumed rate of accretion is much greater than  values assigned to  T Tau disks of ${\sim}$10$^{-8}$ M$_{\odot}$ yr$^{-1}$ \citep{Hartmann2009}.  However, because 10\% of accretion during a star's
pre-main-sequence evolution is accreted during a high luminosity short-lived FU Ori phase \citep{Hartmann2009}, planetesimal formation during this evolutionary phase may occur.  In any case, because Earth's composition is well explained by condensation into solids
at ${\sim}$1100 K \citep{Allegre2001},  a short-lived phase solid-forming when the midplane temperature was near 1400 K at 0.4 AU seems plausible.

Equipped with this simple model, we now extrapolate to  more massive stars to estimate the orbital
location where very refractory planetesimals may form.  Although there is wide variability \citep{Hartmann2009}, we assume  that the relevant accretion rate for the era of the formation of very refractory planetesimals equals our fiducial value of 3 ${\times}$ 10$^{-6}$ M$_{\odot}$ yr$^{-1}$ independent of   $M_{*}$.   Consequently, from Equation (5), we expect that the orbital location of the formation of refractory-dominated planetesimals is:
\begin{equation}
\frac{R_{cond}}{R_{\odot}}\;=\;81\,\left(\frac{M_{*}}{M_{\odot}}\right)^{1/3}\,\left(\frac{{\chi}}{{\chi}_{0}}\right)^{2/9}\,\left(\frac{0.01}{{\alpha}}\right)^{2/9}
\end{equation}
Here,  ${\chi}_{0}$ is the fiducial opacity.

\subsection{Survivability }

Even if very refractory planetesimals formed, they might be engulfed and 
not survive the star's red giant evolution.  We now compare the orbital radius of extremely refractory planetesimals
with the size of the star when it is on the Asymptotic Giant Branch (AGB).

Denote $M_{i}$ and  $M_{f}$ as the initial main-sequence mass and the final white dwarf mass.  The maximum
luminosity on the AGB, $L_{AGB}$, is approximately \citep{Iben1983}:
\begin{equation}
\frac{L_{AGB}}{L_{\odot}}\;{\approx}\;\,5.9\,{\times}\,10^{4}\,\left(\frac{M_{f}}{M_{\odot}}\,-\,0.495\right)
\end{equation}
The initial mass to final mass relationship is \citep{Williams2009}:
\begin{equation}
\frac{M_{f}}{M_{\odot}}\;=\;0.339\;+\;0.129\,\frac{M_{i}}{M_{\odot}}
\end{equation}
For main-sequence stars appreciably more massive than the Sun, then from Equations (9) and (10):
\begin{equation}
\frac{L_{AGB}}{L_{\odot}}\;{\approx}\;7600\,\frac{M_{i}}{M_{\odot}}
\end{equation}
We therefore write for the maximum radius of the star while on the AGB that:
\begin{equation}
\frac{R_{AGB}}{R_{\odot}}\;=\;87\,\left(\frac{M_{i}}{M_{\odot}}\right)^{1/2}\,\left(\frac{T_{\odot}}{T_{AGB}}\right)^{2}
\end{equation}
where $T_{AGB}$ is the star's effective temperature while on the AGB.

With the usual assumption that the planet's orbital radius expands with constant angular momentum as the star loses mass \citep{Veras2011}, then the final distance from the star, $R_{f}$ is:
\begin{equation}
\frac{R_{f}}{R_{cond}}\;=\;\frac{M_{i}}{M_{f}}
\end{equation}
From Equations (8), (12) and (13) equating $M_{i}$ with $M_{*}$, we write that
\begin{equation}
\frac{R_{f}}{R_{AGB}}\;{\approx}\;0.93\,\left(\frac{M_{\odot}}{M_{i}}\right)^{-1/6}\,\left(\frac{T_{AGB}}{T_{\odot}}\right)^{2}\,\left(\frac{{\chi}}{{\chi}_{0}}\right)^{2/9}\left(\frac{0.01}{{\alpha}}\right)^{2/9}\;\left(0.339\;\frac{M_{\odot}}{M_{i}}\;+\;0.129\right)^{-1}
\end{equation}
The requirement that a planetesimal survive into the star's white dwarf evolutionary phase is that $R_{f}$ $>$ $R_{AGB}$. 

\subsection{Results}
We show in Figure 3 a plot of $R_{f}/R_{AGB}$ as a function of a star's initial main-sequence mass.  In all cases, we take $T_{AGB}$ = 3000 K \citep{Bertelli2009}.  For the fiducial values of the viscosity and opacity, we find
that $R_{f}/R_{AGB}$ $<$ 1 for all main-sequence stars of mass less than 6.0 M$_{\odot}$ corresponding to a white dwarf mass of 1.1 M$_{\odot}$.  Therefore we expect that even if  refractory-dominated planetesimals
are formed, unless they are scattered outwards, they are destroyed when the star is on the AGB. However, because the parameters with a protoplanetary disk are uncertain, we also show results in Figure 3 for
a case with higher opacity and/or lower viscosity.  If the opacity is increased by a factor of 10 or
${\alpha}$ is decreased by a factor of 10 -- equivalent changes in Equation (14) -- then refractory-dominated planetesimals might survive through the AGB evolution of main-sequence
for most stars of interest.  Both \citet{Ciesla2010} and \citet{Garaud2007} employ ${\alpha}$ = 0.001, a factor of 10 below our fiducial value; it is possible that this lower value of ${\alpha}$ is appropriate for some protoplanetary disks.   
The results in Figures 1 and 2 provide  support for the fiducial model.

\begin{figure}
 \plotone{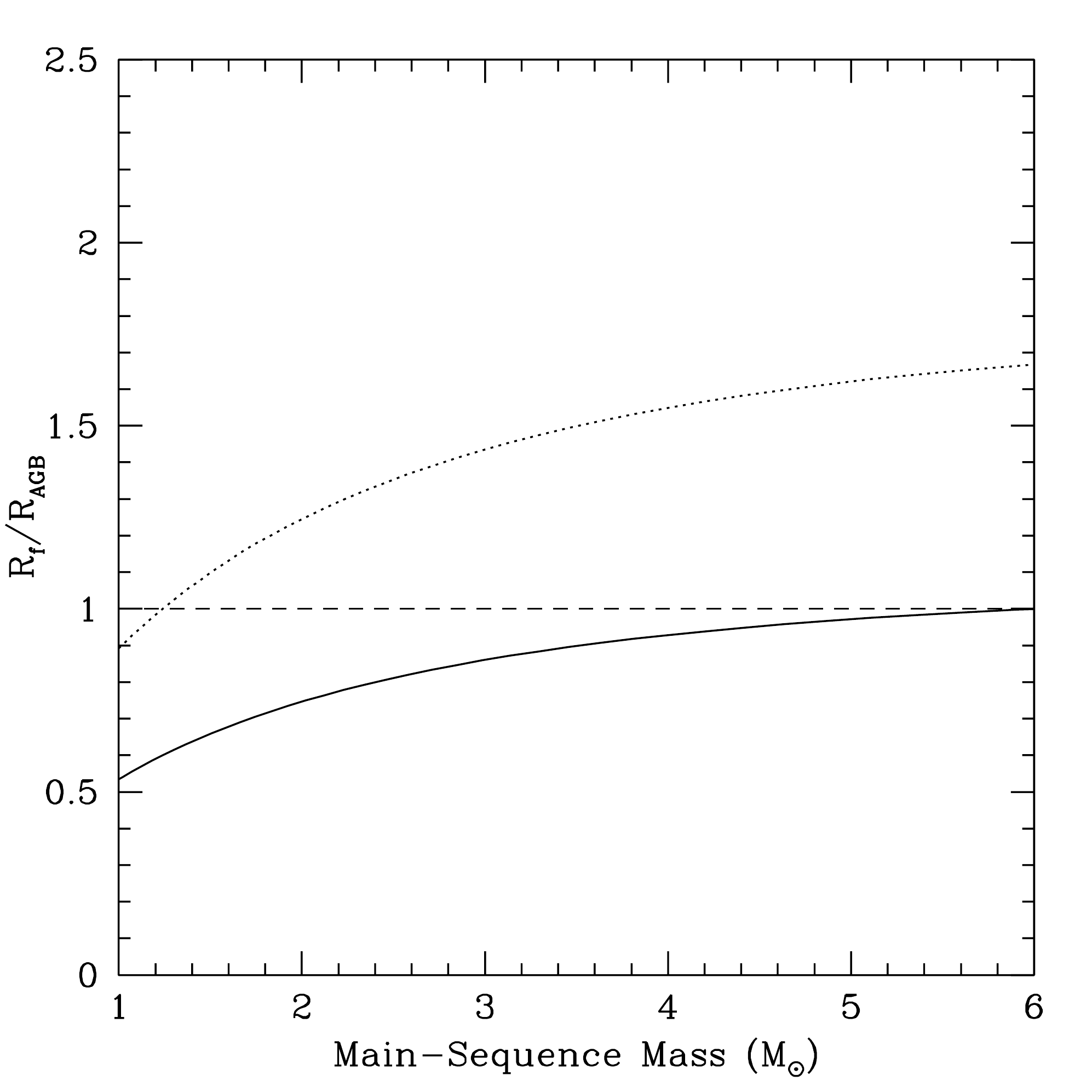}
 \caption{Display of $R_{f}/R_{AGB}$ from Equation (14).     The solid line shows the results
  the fiducial values of the opacity and viscosity.  The dotted line shows the results if the opacity  is
 increased by a factor of 10 or the viscosity parameter, ${\alpha}$ is decreased by a factor of 10.   The horizontal dashed line defines the initial orbital location beyond which  refractory-dominated planetesimals may survive into the star's white dwarf phase.  At least in the fiducial case, exceptionally refractory planetesimals are not expected
 to survive the star's AGB evolution.}
 \end{figure}

\section{DETECTABILITY OF SCATTERED REFRACTORY-DOMINATED PLANETESIMALS}

\cite{Carter-Bond2012} computed that refractory-dominated planetesimals may be dispersed throughout a planetary system by the migration of giant planets.  If so, then
the refractory-dominated planetesimals should survive the star's AGB evolution, and  we may be able to detect them if they are ultimately  accreted onto a white dwarf.  While we see from Figures 1 and 2 that
no known  parent bodies are as refractory-rich as CAIs, perhaps they are subtly masked because their contribution to the observed pollution is diluted.  There are  two possibilities.  First, dispersed refractory-dominated asteroids may have
collided with  ``normal" asteroids and the resulting parent body is only enhanced in refractories but not dominated by them.
Second, perhaps
the observed  white dwarfs have accreted  multiple asteroids.  Because  only a minority of the parent bodies might have compositions dominated by refractories, then we would only detect
an enhancement of refractories but not complete domination.  

Before discussing these possibilities, we consider the entire observed ensemble of polluted white dwarfs.  
 In both Figures 1 and 2, we see that there are Ca-deficient as well as Ca-enhanced
objects.  If the accreted parent bodies were  blends of refractory-enhanced planetesimals with ``normal" objects, there would not be any Ca-deficient asteroids.  
These data therefore favor models where, on average, calcium abundances are determined by  processes in which asteroids differentiate, melt and collide.  In such scenarios, refractory elements can be exchanged among the planetesimal population, but there is no net enhancement.  In contrast, if blending of refractory-dominated
planetesimals into ``normal" asteroids prevailed, then we would never detect objects which are Ca-deficient. Nevertheless, although there is no evidence from the ensemble of examined extrasolar asteroidal material for
the formation of refractory-dominated planetesimals, there may be individual instances where this occurs and below we discuss this possibility.

\subsection{Accretion of Compositionally-blended Parent Bodies?}

  To assess the possibility that there may be particular parent bodies which are composed of a blend of a refractory-dominated planetesimal with a ``normal" object, we consider on a case-by-case basis those white dwarfs where the external pollution is comprehensively measured.  Currently, as listed in Table 1, there are seven polluted dwarfs where at least seven elements heavier than helium, including all four major elements in bulk Earth - O, Mg, Si and Fe,  have been detected and the results published.    Equipped with these data, we can make
a well-constrained estimate of the mass fractions carried in different elements, including  the  refractory elements measured in each star as listed in Table 1. 
 
While the evolutionary history of each individual object is unknown,  a parent body can be considered a candidate for a blend if every detected refractory element  has
a mass fraction significantly larger than its corresponding fraction in bulk Earth.  Because typical uncertainties in abundance determinations are near 0.2 dex and because there is about 0.1 dex dispersion in the relative elemental abundances among main-sequence stars near the Sun, enhancements of an element's mass fraction by less than a factor of two are not clearly
well-established.  
Furthermore, if the accreted parent body is a blend of a refractory-dominated planetesimal with an ambient object, then every refractory
is expected to be enhanced. 
\begin{itemize}
\item{GD 40:  In this star's pollution, the mass fractions of  Ca and Ti are enhanced over bulk Earth's value.  However, according to \citet{Jura2012a}, the  mass fraction of Al in this star's pollution  is 0.013; slightly less than the value of 0.015 for bulk Earth \citep{Allegre2001}.    There is no evidence  for a blend.}
\item{WD J0738+1835:  According to \citet{Dufour2012} the mass fractions of Al, Ca, Ti and Sc are all below Earth's values.  There is no evidence for a blend. }
\item{PG 0843+516:  Al is only 0.005 of the mass fraction of the accreted material \citep{Gaensicke2012}; less than bulk Earth's value of 0.015. There is no evidence for a blend.}
\item{WD 1226+110:  According to \citet{Gaensicke2012}, while elevated, the mass fractions of  Al and Ca are less than a factor of two greater than bulk Earth's.  Consequently, there is at best weak evidence for  blending.}
\item{WD 1929+012:   According to \citet{Gaensicke2012}, the mass fraction of Al is 0.002, much lower than bulk Earth's value.  There is no evidence for a blend.}
\item{G241-6: According to \citet{Jura2012a}, the mass fraction of Al is less than 0.007.  Consequently, there is no evidence for a blend.}
\item{HS 2253+8023: Ti has essentially a solar mass fraction in the measured pollution of this star.  Consequently, there is no evidence for a blend.}
\end{itemize}
At least in this small sample of seven well-studied externally polluted white dwarfs,  there is at best weak evidence that one parent body is a blend of a refractory-dominated planetesimal with 
an ambient asteroid.

\begin{center}
Table 1 --  Summary Evaluation of Possible Blending With a Refractory-Dominated Planetesimal
\\
\begin{tabular}{llrrrrrrrrr}
\hline
\hline
Star &  T$_{eff}$ & Dom. & No. &Refrac. & Dust & Blend &  Ref. \\
\hline
GD 40 &  15,300 & He & 13 & Al, Ca, Ti & Y & N &  (a,b)\\
WD J0738+1835&13,950 & He & 14 & Al, Ca, Ti, Sc &Y & N &  (c) \\
PG 0843+516 &  23,100 & H & 10 &Al & Y & N  &  (d,e) \\
WD 1226+110 & 20,900 & H & 7&  Al, Ca& Y &  ? & (d)\\
WD 1929+012& 21,200 & H & 12 & Al, Ca & Y & N & (d,f,g)\\
G241-6 &  15,300 & He & 12 & Ca, Ti & N & N & (b,h) \\
HS 2253+8023 &  14,400 & He & 9 & Ca, Ti & N & N & (h) \\
\hline
\end{tabular}
\end{center}
The column headings are defined as follows.  ``Dom." reports the dominant light element in the atmosphere. ``No." is the number of elements heavier than helium that are detected.``Refrac" lists the detected refractory elements.  ``Dust"  lists  whether there
is an infrared excess (`Y" = Yes and ``N" = No), ``Blend" gives our estimate of the likelihood for evidence that a refractory-dominated asteroid was blended into
the parent body.   ``Ref" lists references. (a) \citet{Klein2010}; (b) \citet{Jura2012a}; (c) \citet{Dufour2012}; (d) \citet{Gaensicke2012}; (e) \citet{Xu2012};  (f) \citet{Vennes2011}; (g) \citet{Melis2011}; (h) \citet{Klein2011}
\subsection{Single or Multiple Accretion Events?}

\citet{Jura2008} discussed whether we might be witnessing single or multiple accretion events onto a polluted white dwarf.  In this analysis, 
stars with dust disks are exhibiting the tidal disruption of one large parent body.   That is, we would expect different asteroids to have
slightly different orbital inclinations and once their solid material is tidally-disrupted, the mutual collision speeds in the Roche zone between the two sets of dust grains would be
so great that the grains would be vaporized.  Also, if extrasolar asteroids are at all like the solar system's, the bulk of the parent body mass is contained in a relatively
few objects.  It is therefore plausible that the accretion of large amounts of mass is from single objects.  In fact, it is observed that the white dwarfs with the greatest
amounts of pollution are likely to have a dust disk \citep{Kilic2006,Farihi2012b}.

In Figures 1 and 2, we distinguish between stars with dust disks and those without.  We see no evidence that the stars with dust disks -- objects likely  polluted by a single
large parent body -- have a composition on average any different from those perhaps polluted by multiple asteroids.  Therefore, these data do not provide any support for scenarios describing a long-term survival and subsequent accretion of
 intact refractory-dominated planetesimals.

\section{PERSPECTIVE}

Previous work has shown that extrasolar planetesimals
at least qualitatively resemble  rocky planets
in the inner solar system because (i) they are largely composed of O, Mg, Si and Fe; (ii) C is usually deficient by
at least an order of magnitude and (iii) H$_{2}$O usually is less than 1\% of the bulk composition.  These results where volatiles are always deficient can be understood as a result of a condensation sequence in
the protoplanetary nebula \citep{Lee2010,Lodders2003}.  In contrast to the behavior of volatiles which are never seen to be enhanced, as shown in Figures 1 and 2,  refractory elements such as calcium can be either deficient or enhanced.  Therefore, as an ensemble, extrasolar planetesimals
are more than a simple blend of refractory-dominated planetesimals with ``normal" asteroids.   

Further data will be most helpful.  By measuring a large suite of elements in the composition of the parent body, detailed comparison with models becomes possible.  A scenario
of compositional reciprocity where one asteroid's loss is another's gain may explain much of the data for refractory elements in the entire ensemble of extrasolar planetesimals.

 \section{CONCLUSIONS}

We have compiled from the literature measured values
 of $n$(Ca)/$n$(Mg) and $n$(Ca)/$n$(Fe) for externally-polluted white dwarfs; none of the accreted parent bodies are composed primarily  of  refractory material.  We argue on the basis of a toy model for protoplanetary nebula
 that unless dispersed from their original orbital location, refractory-dominated planetesimals likely would be destroyed during the star's AGB evolution.   As yet, there is at best weak evidence
 to support the hypothesis that a refractory-dominated planetesimal was scattered from its zone of formation and then  blended into  material  accreted onto an externally-polluted
 white dwarf.
   
 This work has been partly supported by the NSF.  

\end{CJK}

 \bibliographystyle{apj}

\end{document}